\documentclass{article}

\PassOptionsToPackage{numbers, sort&compress}{natbib}

\usepackage{amsmath} 

\usepackage[final]{neurips_2024_ml4ps}

\usepackage[utf8]{inputenc} 
\usepackage[T1]{fontenc}    
\usepackage{hyperref}       
\usepackage{url}            
\usepackage{booktabs}       
\usepackage{amsfonts}       
\usepackage{nicefrac}       
\usepackage{microtype}      
\usepackage{xcolor}         

\usepackage{wrapfig}
\usepackage{graphicx} 		
\usepackage{tikz}

\newcommand{\hMpc}{{\ifmmode{h^{-1}{\rm Mpc}}\else{$h^{-1}$Mpc}\fi}}
\newcommand{\Mpc}{{\ifmmode{{\rm Mpc}}\else{Mpc}\fi}}
\newcommand{\hkpc}{{\ifmmode{h^{-1}{\rm kpc}}\else{$h^{-1}$kpc}\fi}}
\newcommand{\kpc}{{\ifmmode{ {\rm kpc} }\else{{\rm kpc}}\fi}}
\newcommand{\kms}{{\ifmmode{ {\rm km\,s^{-1}} }\else{ ${\rm km\,s^{-1}}$ }\fi}}
\newcommand{\hMsun}{{\ifmmode{h^{-1}{\rm {M_{\odot}}}}\else{$h^{-1}{\rm{M_{\odot}}}$}\fi}}
\newcommand{\Msun}{{\ifmmode{{\rm M}_{\odot}}\else{${\rm M}_{\odot}$}\fi}}
\newcommand{\Mhalo}{{\ifmmode{M_{\rm halo}}\else{$M_{\rm halo}$}\fi}}
\newcommand{\Rvir}{{\ifmmode{R_{\rm vir}}\else{$R_{\rm vir}$}\fi}}
\newcommand{\Mvir}{{\ifmmode{M_{\rm vir}}\else{$M_{\rm vir}$}\fi}}
\newcommand{\Mstar}{{\ifmmode{M_{\rm star}}\else{$M_{\rm star}$}\fi}}
\newcommand{\Vrot}{{\ifmmode{V_{\rm rot}}\else{$V_{\rm rot}$}\fi}}
\newcommand{\ltsima}{$\; \buildrel < \over \sim \;$}
\newcommand{\gtsima}{$\; \buildrel > \over \sim \;$}
\newcommand{\lsim}{\lower.5ex\hbox{\ltsima}}
\newcommand{\gsim}{\lower.5ex\hbox{\gtsima}}

\def\lesssim{\mathrel{\hbox{\rlap{\hbox{\lower4pt\hbox{$\sim$}}}\hbox{$<$}}}}
\def\gtrsim{\mathrel{\hbox{\rlap{\hbox{\lower4pt\hbox{$\sim$}}}\hbox{$>$}}}}

\newcommand{\beq}{\begin{equation}}
\newcommand{\eeq}{\end{equation}}
\def\beqa{\begin{eqnarray}}
\def\eeqa{\end{eqnarray}}
\def\LCDM{\ensuremath{\Lambda}CDM}

\def \kms {\ifmmode  \,\rm km\,s^{-1} \else $\,\rm km\,s^{-1}  $ \fi }
\def \kpc {\ifmmode  {\rm kpc}  \else ${\rm  kpc}$ \fi  }  
\def \hkpc {\ifmmode  {h^{-1}\rm kpc}  \else ${h^{-1}\rm kpc}$ \fi  }  
\def \hMpc {\ifmmode  {h^{-1}\rm Mpc}  \else ${h^{-1}\rm Mpc}$ \fi  }  
\def \Mpch {\ifmmode  {h^{-1}\rm Mpc}  \else ${h^{-1}\rm Mpc}$ \fi  }  
\def \Msun {\ifmmode {\rm M}_{\odot} \else ${\rm M}_{\odot}$ \fi} 
\def \hMsun {\ifmmode h^{-1}\,\rm M_{\odot} \else $h^{-1}\,\rm M_{\odot}$ \fi}

\def \LCDM {\ifmmode \Lambda{\rm CDM} \else $\Lambda{\rm CDM}$ \fi}
\def \sig8 {\ifmmode \sigma_8 \else $\sigma_8$ \fi} 
\def \OmegaM {\ifmmode \Omega_{\rm m} \else $\Omega_{\rm m}$ \fi} 
\def \Omegab {\ifmmode \Omega_{\rm b} \else $\Omega_{\rm b}$ \fi} 
\def \OmegaL {\ifmmode \Omega_{\rm \Lambda} \else $\Omega_{\rm \Lambda}$\fi} 
\def \Deltavir {\ifmmode \Delta_{\rm vir} \else $\Delta_{\rm vir}$ \fi}
\def \rhocrit {\ifmmode \rho_{\rm crit} \else $\rho_{\rm crit}$ \fi}
\def \rhou {\ifmmode \rho_{\rm u} \else $\rho_{\rm u}$ \fi}
\def \zc {\ifmmode z_{\rm c} \else $z_{\rm c}$ \fi}

\title{Differentiable Conservative Radially Symmetric Fluid Simulations and Stellar Winds $\circ$ jf1uids}

\author{%
  Leonard Storcks {\rm and} Tobias Buck \\
  Interdisciplinary Center for Scientific Computing, University of Heidelberg \\
  Im Neuenheimer Feld 205, D-69120 Heidelberg, Germany \\
  \href{mailto:leonard.storcks@stud.uni-heidelberg.de}{\texttt{leonard.storcks@stud.uni-heidelberg.de}} \& \href{mailto:tobias.buck@iwr.uni-heidelberg.de}{\texttt{tobias.buck@iwr.uni-heidelberg.de}}
}

\begin{document}

\maketitle

\begin{abstract}
We present \texttt{jf1uids}, a one-dimensional fluid solver that can, by virtue of
a \textit{geometric formulation} of the Euler equations, model radially symmetric fluid problems in a conservative manner, i.e., without losing
mass or energy. For spherical problems, such as ideal supernova explosions or stellar
wind-blown bubble expansions, simulating only along a radial dimension drastically
reduces compute and memory demands compared to a full three-dimensional method. This simplification 
also alleviates constraints on backpropagation through the solver. Written in \texttt{JAX}, \texttt{jf1uids} is a GPU-compatible and fully differentiable simulator. We demonstrate the advantages of this differentiable physics simulator by retrieving the wind's parameters for an adiabatic stellar wind expansion  from the final fluid state using gradient descent.
As part of a larger "stellar winds, cosmic rays and machine learning" research track, \texttt{jf1uids} serves as a solid foundation to be extended with additional physics modules, foremost
cosmic rays and a neural-net powered gas-cooling surrogate and improved by higher order and more accurate numerical schemes.
All code is available under \href{https://github.com/leo1200/jf1uids/}{https://github.com/leo1200/jf1uids/}.
\end{abstract}

\section{Motivation and related work}

\paragraph{Our motivation is blowin' in the wind} We are interested in detailed simulations of radial gaseous outflows from a star (stellar winds) \cite{weaver, parker58} and ultimately, inspired by recent observations \cite{Peron2024}, their interaction with cosmic rays (energetic particles), a challenging pursuit in multiple ways. High-resolution three-dimensional simulations are computationally expensive, and physical effects like (turbulent) cooling of the gas in practice often scale with the resolution of the simulation (dependent on how well the bubble surface is resolved) \cite{lachlan21,lachlan24}. We therefore aim for a performant simulation code with radial symmetry,\footnote{This might also be of interest in other areas like studying sonoluminescent bubbles \cite{barber}.} in which detailed physics is modeled by machine learning surrogates in the solver. Finally, once detailed physics is incorporated, it would be desirable to use the simulator to reconstruct stellar wind parameters from observed data (inverse modeling).
\paragraph{Towards high-performance} Low level classical simulation codes (like \cite{Springel10, Stone08}) with optimized hand-crafted parallelization schemes are in the process of being extended to High-Performance Computing (HPC) on GPUs (and TPUs) (see \cite{zier24, grete20}) based on vendor specific GPU-programming language extensions like CUDA \cite{cuda} or a framework like \texttt{kokkos} \cite{kokkos}. In another avenue for HPC, which we follow, most of the underlying complexity is abstracted into higher-level numerical libraries like \texttt{tinygrad} \cite{tinygrad2024}, \texttt{PyTorch} \cite{pytorch} or \texttt{JAX} \cite{jax} in the latter of which various simulation codes have already been written (e.g. \cite{pmwd,jaxcosmo, diffrax, jax_fem, jax_sph, jax_fluids}).\footnote{Note, however, that adaptive multiresolution (AMR) in a differentiable and just-in-time compiled domain-specific language like JAX has not yet been achieved \cite{jaxfluids2} but is generally possible in a HPC setting (see \cite{parthenon}). We provide a simple prototype for AMR in one dimension in \texttt{JAX} under \href{https://github.com/leo1200/jamr}{https://github.com/leo1200/jamr}.}
\paragraph{Differentiability} Another advantage is, that code written in JAX is differentiable, the result of a whole simulation can be differentiated with respect to its parameters. This allows us to directly optimize machine learning surrogates for the detailed physics in the simulator, rather than learning an error correction scheme (like NeuralSim \cite{neuralsim} does in the field of robotics). The same goes for optimizing parts of the
numerics like gradient limiters \cite{deep_flux_limiters}, or smoothness indicators \cite{deep_smoothness_indicators}. Inverse modeling also benefits from differentiability - in the simplest cases, direct gradient descent on the parameters may be feasible.

\paragraph{Why not JAX-Fluids?} \texttt{JAX-Fluids}\footnote{\href{https://github.com/tumaer/JAX-Fluids}{https://github.com/tumaer/JAX-Fluids}} is a fluid simulator written in \texttt{JAX} featuring high-order\footnote{The spatial order of a fluid simulator relates to the speed of numerical convergence with increasing grid resolution (in smooth regimes).} numerical schemes. While it supports one-dimensional spherically symmetric simulations, it does so via a geometric source formulation \cite[section 1.6.4]{Toro2009} which is not conservative. Total energy and mass are not constant over time, an unwanted effect \cite{euler_spherical}.

\section{The numerics of jf1uids}
\subsection{Geometric source term vs. geometric formulation of the Euler equations}
The geometric difference of a radially symmetric two- or three-dimensional to a Cartesian one-dimensional inviscid fluid can be encompassed by a geometric source term $\vec{S}(\vec{U})$ in the Euler equations
\begin{equation}
    \partial_t \vec{U} + \partial_r \vec{F} = \vec{S}
\end{equation}
where the state vector $\vec{U}$, the flux vector $\vec{F}$ and geometric source term $\vec{S}$ are given by
\begin{equation}
\vec{U}=\left[\begin{array}{c}
\rho \\
\rho u \\
E
\end{array}\right], \quad \vec{F}=\left[\begin{array}{c}
\rho u \\
\rho u^2+p \\
u(E+p)
\end{array}\right], \quad \vec{S}=-\frac{\alpha}{r}\left[\begin{array}{c}
\rho u \\
\rho u^2 \\
u(E+p)
\end{array}\right], \quad \alpha = \begin{cases} 
0 & \text{Cartesian} \\
1 & \text{Cylindrical} \\
2 & \text{Spherical}
\end{cases}
\end{equation}
with density $\rho$, velocity $u$, energy $E$ and pressure $p$ given via an equation of state. In this formulation, expanding a standard Cartesian one-dimensional solver to radially symmetric two- and three-dimensional simulations only requires implementing the source term $\vec{S}$. This was done in \texttt{JAX-Fluids} source term solver. Only adapting the solver in terms of a source term, however, leads to numerics inconsistent with the geometry - the proper cell centers to use are the volumetric not the geometric centers and for fluxes to be conservative, the radial increase of area the flux passes through is to be considered (more formal reasoning can be found in \cite{euler_spherical}).

Based on the geometric form of the Euler equations
\begin{equation}
    \partial_t \vec{U} +\frac{1}{r^\alpha} \partial_r (r^\alpha \vec{F}) = \frac{1}{r^\alpha}\vec{N}, \quad \text{with pressure nozzling } \vec{N} = \left[\begin{array}{c}
0 \\
\alpha r^{\alpha-1}p \\
0
\end{array}\right],
\end{equation}
\citet{euler_spherical} present the baseline discretization for a radial fluid solver which we implement in Hancock's second order MUSCL-scheme (as described in \cite{vanLeer2003}; we opt for the \textit{minmod gradient limiter}) with a HLL Riemann solver \cite{Toro1997}.

\subsection{On conservational properties}
\begin{figure}[!htb]
  \centering
  \includegraphics[width=1.0\textwidth]{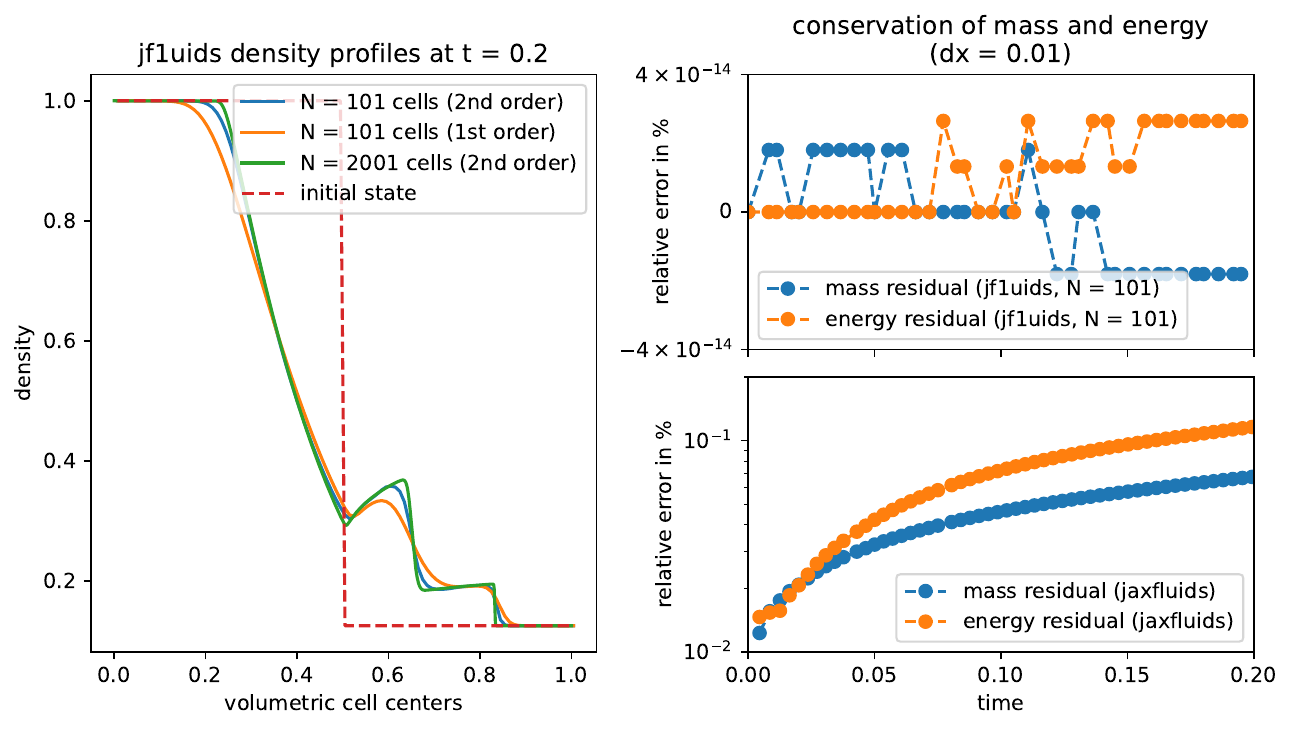}\hfill
  \caption{On the left side, the illustrative problem, a radial shock setting is presented, and different solutions compared, on the right the change in mass and energy over time for \texttt{jf1uids} and \texttt{JAX-Fluids} is shown.}
  \label{fig:shock_problem}
 \end{figure}

To give an example of simulation results obtained using \texttt{jf1uids} we depict the final density over the radial dimension in a shock test setting with step-functions as initial conditions for the density and pressure at $t = 0.2$ on the left side of Fig. \ref{fig:shock_problem}. For different resolutions and different spatial orders (first order Godunov and second order MUSCL-Hancock scheme (see \cite{Toro2009} for details)), as expected, higher resolutions and higher-order spatial schemes lead to sharper solutions.

\texttt{jf1uids} preserves mass and energy throughout a fluid simulation (for a proof, see \cite{euler_spherical}). As noted by \citet{euler_spherical}, this is a consequence of the discretization of the geometric form of the Euler equations used, not the specific flux-calculation method, the spatial order or the grid size etc. We demonstrate mass and energy conservation for the shock problem in Fig. \ref{fig:shock_problem}.

While \texttt{jf1uids} energy and mass residuals fluctuate between $\pm4\cdot 10^{-14} \%$,\footnote{This is due to the limited precision in which the calculations are carried out. Especially with single precision floating point calculations, slight drifts in the \textit{conserved} quantities are possible (just as in symplectic ordinary differential equation solvers, see \cite{ias15}).} \texttt{JAX-Fluids} shows increasing residuals with a relative energy error up to $0.1\%$. Note that the same conservative baseline discretization could - with some implementation effort - also be used with the higher-order methods of \texttt{JAX-Fluids} (see also \cite{wang17}).

\subsection{On performance}
\texttt{jf1uids} is written in pure \texttt{JAX}, can be just-in-time compiled and therefore runs on CPUs, GPUs and TPUs alike. Detailed performance-accuracy comparisons between \texttt{jf1uids} and \texttt{JAX-Fluids} will be provided in the future but are not straightforward. Higher order methods provide better accuracy at lower resolution but increased cost, such that, depending on the setting and exact methods, the accuracy per compute of lower order methods can be superior (as in \cite{order_comparison}).

\section{Differentiation through stellar wind simulations}

\begin{figure}[!htb]
  \centering
  \includegraphics[width=1.0\textwidth]{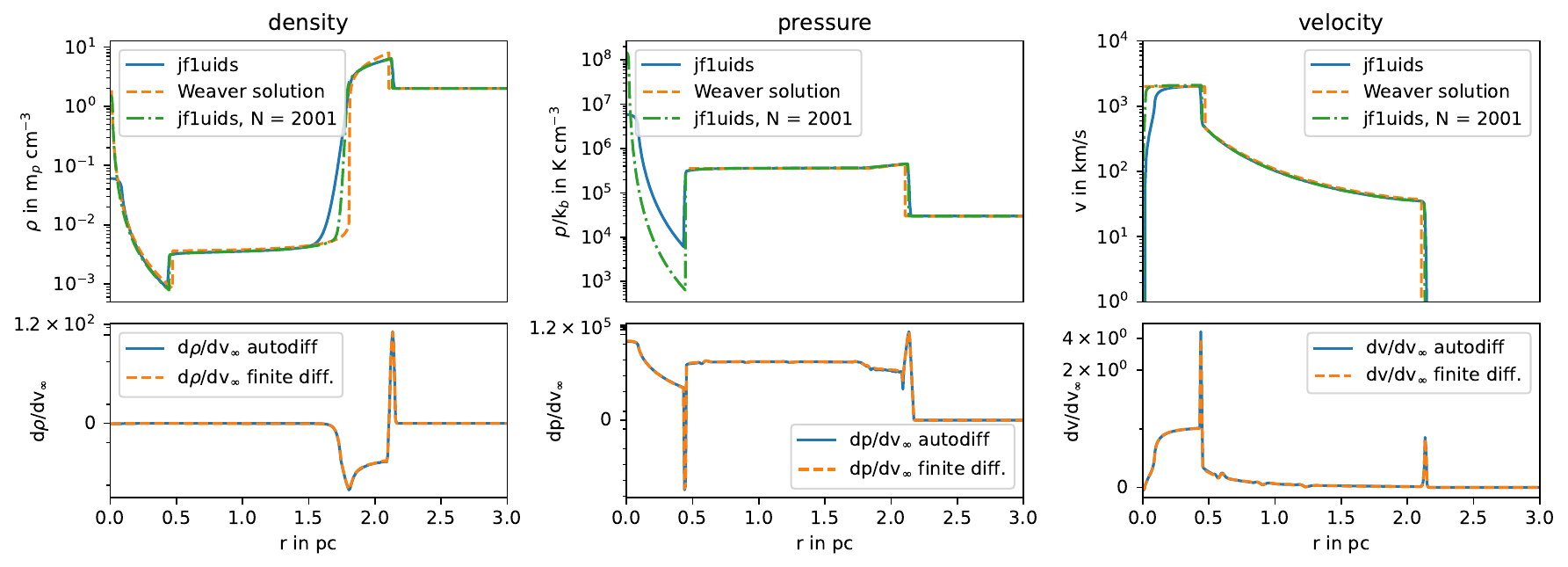}\hfill
  
  \vspace{-.25cm
  }
  \caption{Density, pressure and velocity of a stellar wind simulation in \texttt{jf1uids} and their derivatives with respect to the outflow velocity of the stellar wind $v_\infty$ at $v_\infty = 2000 \frac{\text{km}}{\text{s}}$. The resolution is $N=401$ cells. The automatic differentiation (autodiff) was done with \texttt{jacfwd}, the finite difference plotted is a centered difference quotient with $\Delta v_\infty = 0.1 \frac{\text{km}}{\text{s}}$}
  \label{fig:weaver_sensitivity}
 \end{figure}

As a first physics module, we have implemented stellar wind based on the injection of thermal energy.\footnote{Thermal energy injection is comparable to the situation in a small cluster of stars with thermalization of kinetic energy \cite{lachlan21}, for a comparison of injection schemes, see \cite{pittard21}.} Our simulation successfully replicates the theoretical Weaver solution \cite{weaver} (see Fig. \ref{fig:weaver_sensitivity}). The main parameters of the wind are the stellar mass $M$ and outflow velocity of the wind $v_\infty$. Fig. \ref{fig:weaver_sensitivity} shows the simulation of an exemplary wind and the derivative of the final fluid state with respect to $v_\infty$ - \texttt{jf1uids} is fully differentiable. The derivatives are remarkably stable and have simple physical interpretations. For example, nudging up $v_\infty$ will forward-displace the shock, leading to the spike in the derivatives after $2$ parsec. Note that for backpropagation through the solver, all the internal operations (so the whole trajectory over time, or part of it based on checkpointing \cite{jaxfluids2}) are stored \cite{diffrax} which can become a memory bottleneck. This, however, is less pronounced in our one-dimensional compared to a full three-dimensional simulation.\footnote{At high resolutions, which also necessitate smaller time steps \cite{CFL}, this can still be an issue.}

\begin{figure}[!htb]
  \centering
  \includegraphics[width=1.0\textwidth]{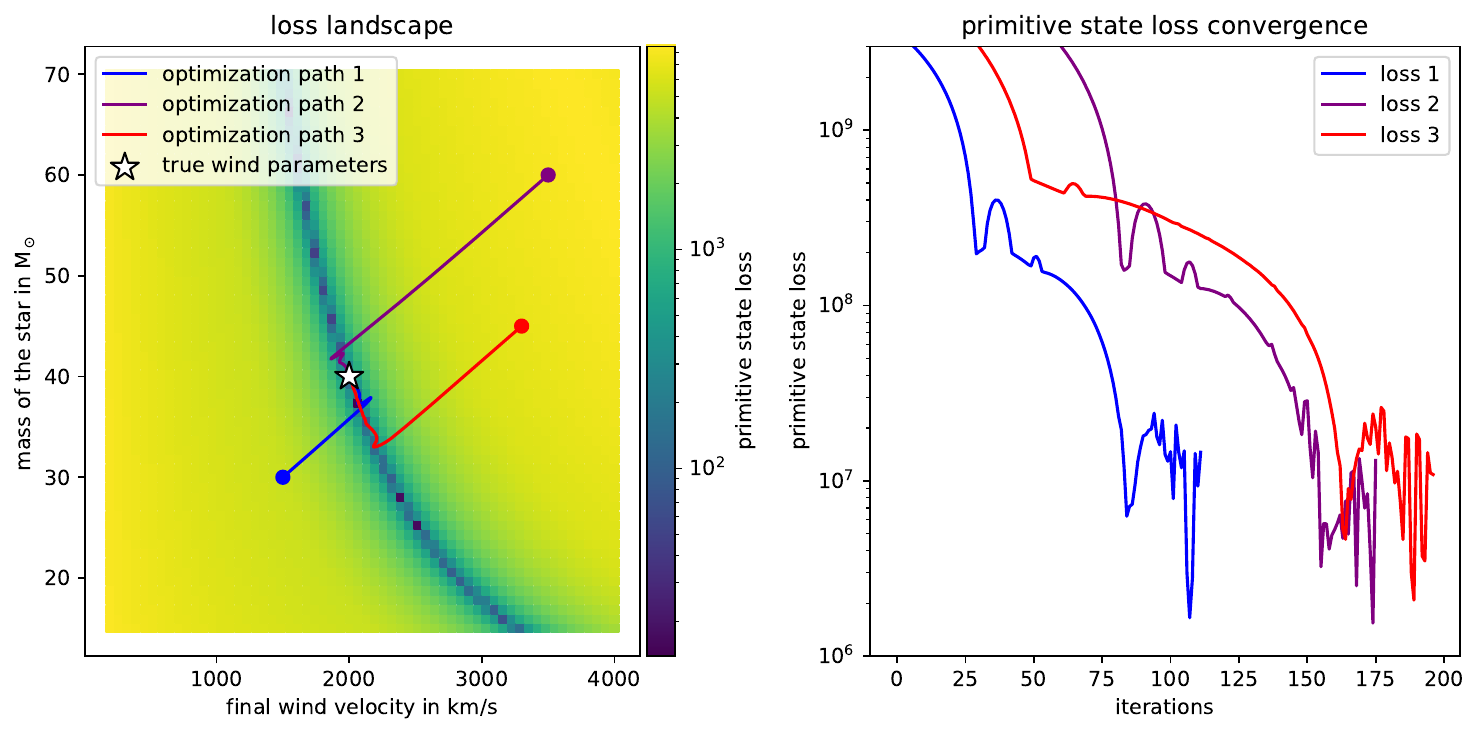}\hfill
  
  \vspace{-.25cm
  }
  \caption{Gradient descent for retrieving stellar wind parameters from the final fluid state of a stellar wind simulation. On the left side, the optimization paths through the two-dimensional parameter space of the wind are shown, on the right, a loss measuring the summed absolute difference to the final state we seek to attain is shown. The resolution is $N=401$ cells.}
  \label{fig:optimization}
 \end{figure}

 In Fig. \ref{fig:optimization} we further show, that based on the full final fluid state, simulated using \texttt{jf1uids}, and the general simulation setup (background density, pressure, adiabatic index, ...) by gradient descent through the simulation, we can reconstruct stellar wind parameters.

\section{Conclusion and outlook}
\begin{figure}[!htb]
  \centering
  \includegraphics[width=1.0\textwidth]{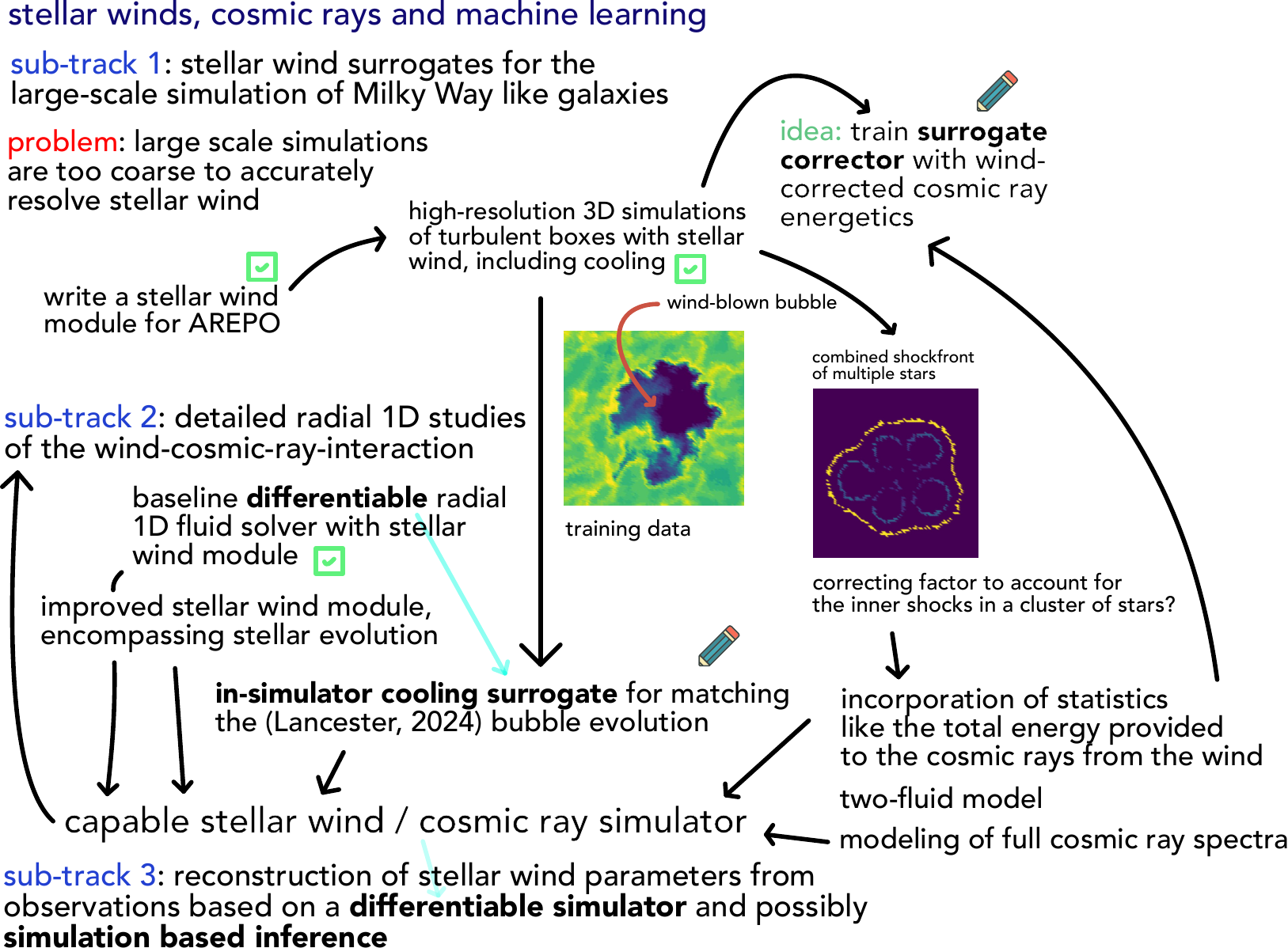}\hfill
  \caption{Overview on the broader research context of this work.}
  \label{fig:researchtrack}
 \end{figure}

We have presented a fully differentiable, conservative and GPU-ready radial fluid solver in \texttt{JAX}, with superior conservative properties compared to \texttt{JAX-Fluids} \cite{jaxfluids2}. Differentiability was showcased on stellar wind simulations. Further research as part of the "stellar winds, cosmic rays and machine learning" research track aims at (also see the research overview illustration Fig. \ref{fig:researchtrack})
\enlargethispage{\baselineskip}
\begin{itemize}
    \item \textit{numerics:} implementation of higher-order methods and more advanced Riemann solvers (lack thereof is a current limitation), detailed performance profiling
    \item \textit{machine learning enhanced numerics:} end-to-end training of gradient limiters and smoothness indicators to improve upon the baseline numerics
    \item \textit{machine learning powered physics:} surrogate modeling for cooling; based on analytical models with free parameters and neural-network cooling terms; aiming for wind evolutions as theorized in \cite{lachlan24}
    \item \textit{cosmic rays:} from summary statistics based on shock energy dissipation \cite{cr_simple} over two-fluids models \cite{cr_two_fluids} to full cosmic ray energy distributions co-modeled with the fluid flow \cite{cr_full}; the dependence of cosmic ray acceleration on the evolution of the wind will be investigated
    \item \textit{parameter retrieval}: insights into stellar wind parameters from observable summary statistics via gradient descent, Hamilton-Monte-Carlo \cite{hmc} and simulation-based inference \cite{sbi}, taking advantage of the differentiability of the simulator \cite{zeghal22}
\end{itemize}

\newpage

\section*{Broader impact statement}
The authors are not aware of any immediate ethical or societal implications of this work. This work purely aims to aid scientific research and proposes a method for radial fluid simulations.

\bibliography{references}

\end{document}